**Title:** A Comparative study of Artificial Neural Networks Using Reinforcement learning and Multidimensional Bayesian Classification Using Parzen Density Estimation for Identification of GC-EIMS Spectra of Partially Methylated Alditol Acetates on the World Wide Web

**Authors:** Faramarz Valafar, Homayoun Valafar

**Conference:** IC-AI '99

**Paper identification number:** 265A

# A Comparative study of Artificial Neural Networks Using Reinforcement learning and Multidimensional Bayesian Classification Using Parzen Density Estimation for Identification of GC-EIMS Spectra of Partially Methylated Alditol Acetates on the World Wide Web

Faramarz Valafar
University of Georgia, CCRC
Athens, GA 30602
faramarz@ccrc.uga.edu

Homayoun Valafar
University of Georgia, CCRC
Athens, GA 30602
homayoun@ccrc.uga.edu

*Abstract:*

*This study reports the development of a pattern recognition search engine for a World Wide Web-based database of gas chromatography-electron impact mass spectra (GC-EIMS) of partially methylated Alditol Acetates (PMAAs). Here, we also report comparative results for two pattern recognition techniques that were employed for this study. The first technique is a statistical technique using Bayesian classifiers and Parzen density estimators. The second technique involves an artificial neural network module trained with reinforcement learning. We demonstrate here that both systems perform well in identifying spectra with small amounts of noise. Both system's performance degrades with degrading signal-to-noise ratio (SNR). When dealing with partial spectra (missing data), the artificial neural network system performs better. The developed system is implemented on the world wide web, and is intended to identify PMAAs using submitted spectra of these molecules recorded on any GC-EIMS instrument. The system, therefore, is insensitive to instrument and column dependent variations in GC-EIMS spectra.*

## 1. Introduction:

**GC-EIMS spectra of PMAAs.** An example of a relatively clean GC-EIMS [1-3] spectrum of a PMAA [4-5] is shown in figure 1 (located at the end of the article). These spectra normally suffer from such aberrations as chemical and instrumental noise, varying signal strength, and varying relative peak intensities. For instance, if the peaks in the GC trace do not resolve well, the spectrum can be a mixture of GC-EIMS spectra of two or more PMAAs. The same effect can be observed, if some of the compound from previous GC-EIMS run is left behind in the column and becomes mixed with the current compound. Furthermore, some instrument operators do not record the entire mass per charge (m/z) range (50-350 m/z) shown in figure 1 (see figure 2 at the end of the article). This could be due to the fact that the operator has some idea as to the type of the PMAA, and for instance decides that only recording and comparing a subset of the mentioned range would be sufficient for the identification of the compound. In this case, an automated recognition (identification) system for these compounds must also be trained to handle missing data.

## 2. Method:

Two pattern recognition systems were used to build an instrument-independent, noise tolerant identification system. The first system uses multidimensional Bayes' classification [6] that uses Parzen density estimation [7] for estimating the *a priori* probability density functions required by the Bayesian classifier. The second system uses a feed-forward, 2-stage artificial neural network (ANN) [8] trained with back-propagation learning algorithm [8] in combination with reinforcement learning [9]. Both systems were trained with 90 spectra of 45

PMAAs, and tested with 385 newly recorded spectra of the same PMAAs (or their configurational isomers [10]). Each spectrum contained 301 points representing the region 30-350 m/z (mass per unit charge). Five percent normally distributed noise was dynamically added to the spectra at the beginning of each training epoch to prevent memorization. The optimal network configuration for the ANN model was found to be 301 input, 20 hidden, and 45 output neurons. The standard grid method was employed for the reinforcement learning [9, 11].

## 3. Results:

The performance of the ANN was compared to that of a multidimensional Bayesian classifier. Parzen density estimation was used to estimate the multidimensional probability distribution functions required by the Bayes' classifier.

Table 1, illustrates the results of the training and testing sessions for both methods. The ANN model learned all the patterns in the training set, and performed equally well on the testing patters that it had not observed during the training session. The statistical model had difficulty learning three of the ninety (or 3.3%) training patterns. Correspondingly, it misclassified 16 of the 385 (or 4.16%) of the testing patterns. Although both systems performed with a high accuracy, the statistical model left some room for improvement.

As mentioned before five percent normally distributed noise was added to the patterns during training. The results shown for the training session are therefore the results of the two methods learning noisy spectra. The testing results reported in Table 1, however, are the results of submitting the testing set to the model without any addition of any noise.

Table 1. Comparative performances of ANN and Bayes' Classifier.

| Classification Method | Training Set | Testing Set |
|---|---|---|
| Parzen density estimation / Bayesian Classification | 96.67% | 95.84% |
| Artificial Neural Network- Reinforcement learning | 100% | 100% |

In these experiments ±5% and ±10% normally distributed noise was added to the testing set creating three new testing sets. The two trained modules were tested with the three noisy testing sets. Table 2, reports the results of these experiments.

Table 2. Comparative performances of ANN and Bayes' Classifier in presence of noise.

| Classification Method | Testing Set ±5% noise | Testing Set ±10% noise |
|---|---|---|
| Parzen density estimation / Bayesian Classification | 96.62% | 90.91% |
| Artificial Neural Network- Reinforcement learning | 100% | 91.43% |

As it can be seen, the performance of neither models degraded significantly when tested with the ±5% noise testing set. In fact, the performance of the statistical model improved to about the same level (3.38% error, or 13 of 385 patterns misclassified) as its performance with the training set. We speculate that this is due to the fact that we built both models with ±5% noise added to the spectra. It is therefore possible that the models, especially the statistical model, has learned some characteristics of the noise and performs better when those characteristics are present. The ANN model's performance degraded from 100% to 99.74%, or 1 out of 385 patterns misclassified. The performance of both models degraded significantly as the noise level was increased to ±10%. The performance of the statistical model degraded to 90.91%, or 35 of 385 patterns misclassified, while the performance of the ANN model degraded to 91.43%, or 33 of 385 patterns misclassified. We further speculate that the performance of both models continue to decrease with the decreasing SNR. Table 3, offers performance statistics of both models when presented with partial spectra.

Table 3. Comparative performances of ANN and Bayes' Classifier in missing data cases. The range 50-90 m/z was deleted from the testing spectra.

| Classification Method | 90-350 m/z spectra |
|---|---|
| Parzen density estimation / Bayesian Classification | 62.08% |
| Artificial Neural Network- Reinforcement learning | 78.18% |

In this experiment we eliminated the peak information bellow 90 m/z for all the test spectra. Here, a distinct difference between the performance of the two models was noticeable. The ANN model seemed to be able to deal with missing data better than the Bayes' model. While Bayes' model's performance decreased to 62.08%, or 146 of the 385 patterns were misclassified, the ANN model misclassified 84 of the 385 patterns with an accuracy of 78.18%. This accounts for over 16% higher accuracy for the ANN model compared to the statistical model.

## 4. New Aspects:

Separation of PMAAs based on their GC-EIMS spectra recorded on any instrument or column is a non-linearly separable task. To the best knowledge of the authors, this is the first identification system for this group of molecules that is instrument/column insensitive and uses ANNs. The system has been implemented on the web, and already has been widely used by scientists around the world to correctly identify over 1,500 spectra at http://www.ccrc.uga.edu/web/ccrcnet/ParsKimi.htm.

## 5. Conclusions:

PMAAs are monosaccharide structures and are the building blocks of all complex carbohydrates. They can be found in all plant and animal life forms. Rapid identification of these molecules are highly desirable. While it is relatively strait forward to build an instrument dependent statistical or rule based (peak matching) model that will identify these molecules, the building of an instrument / column independent statistical model to identify these molecules has proven to be difficult. We have developed an artificial neural network system that can successfully distinguish between the GC-EIMS spectra of these molecules. Our comparative study shows that ANNs in combination with reinforcement learning perform slightly better than Bayes' multidimensional classifiers using Parzen density estimators. We have further shown that the system can be trained to be noise (chemical and instrumental) insensitive to a reasonable degree. Finally, our comparative study indicates that ANN are better in handling missing data than Bayes' classifiers in the problem of instrument/column independent identification of GC-EIMS spectra of PMAAs.

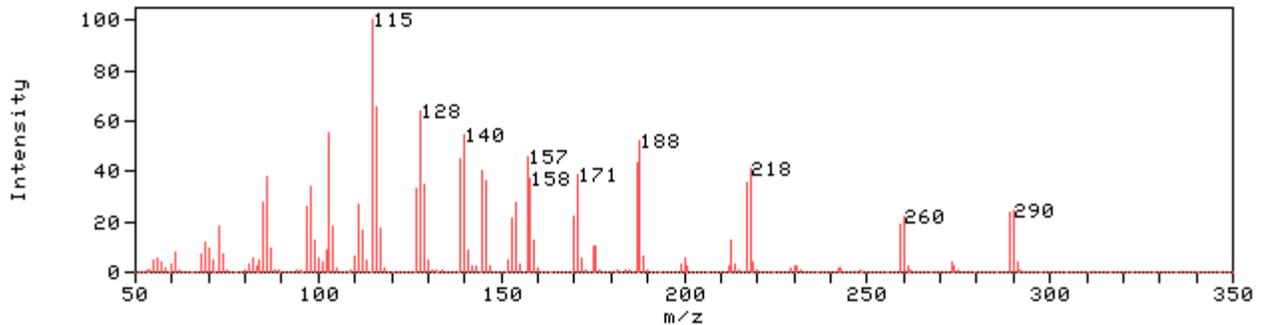
**Figure 1.** A GC-EIMS spectrum of a 2,3,4,6-linked Hexopyranose (Galactose).

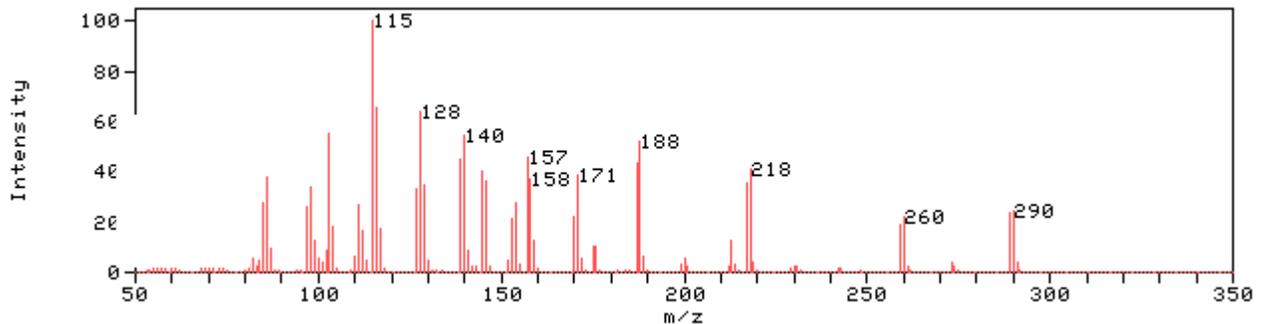
**Figure 2.** A partial GC-EIMS spectrum of a 2,3,4,6-linked Hexopyranose (Galactose). Information from 50 to 90 m/z was intentionally deleted.